\documentclass[a4paper, aps, prx,superscriptaddress,twocolumn,a4paper]{revtex4-2}
\usepackage{bm}
\usepackage{graphicx}

\usepackage{amsmath}
\usepackage{amssymb}%
\usepackage{color}

\usepackage[colorlinks=true,citecolor=blue,urlcolor=blue]{hyperref}
\usepackage{nameref}

\begin{document}
\title{Quantum enhancement and Doppler suppression of Kasevich-Chu atom interferometer with motional squeezing states}	
\begin{abstract}
	Hybridization of internal and external atomic degrees of freedom in a Kasevich-Chu interferometer enables the possibility to enhance the sensitivity significantly 
	even under quantum-standard limit. 
	By introducing motional squeezing state as an input, 
	we systematically derive the computational framework of quantum and classical Fisher information of two measurement protocols for arbitrary strength of Doppler effects. 
	Through maximizing the corresponding classical Fisher information, we obtain the optimal control parameters and the corresponding quantum Fisher information. 
	For population measurement, the largest sensitivity can be as large as four times than the semi-classical limit through enlarging the atom coherence length. 
	For joint measurement of population and position, 
	the competition between quantum enhancement and Doppler suppression induces two three behaviors, 
	in one regime, the quantum enhancement dominates even in presence of strong Doppler broadening effects where the sensitivity is significantly enhanced; while in another regime, an optimal squeezing parameter is observed where the classical Fisher information reaches the maximum. 
	Our results clearly demonstrate the robustness of external quantum enhancement against Doppler suppression. 
	Our proposal can be readily applied to gravimeter of mobile platform where decoherence from noise will damage the many-body entanglement of internal spin squeezing.
\end{abstract}
\date{\today}
\author{Dongyang Yu}
\affiliation{School of Physics and Optical Engineering, Zhejiang University of Technology, No. 288, Liuhe Road, Hangzhou, 310023, China}
\affiliation{Zhejiang Provincial Key Laboratory and Collaborative Innovation Center for Quantum Precise Measurement, Zhejiang University of Technology, No. 288, Liuhe Road, Hangzhou, 310023,  China}
\author{Yubin Wang}
\affiliation{School of Physics and Optical Engineering, Zhejiang University of Technology, No. 288, Liuhe Road, Hangzhou, 310023, China}
\affiliation{Zhejiang Provincial Key Laboratory and Collaborative Innovation Center for Quantum Precise Measurement, Zhejiang University of Technology, No. 288, Liuhe Road, Hangzhou, 310023,  China}
\author{Feng en Oon}
\affiliation{School of Physics and Optical Engineering, Zhejiang University of Technology, No. 288, Liuhe Road, Hangzhou, 310023, China}
\affiliation{Zhejiang Provincial Key Laboratory and Collaborative Innovation Center for Quantum Precise Measurement, Zhejiang University of Technology, No. 288, Liuhe Road, Hangzhou, 310023,  China}
\author{Qiang Lin}
\affiliation{State Key Laboratory of Ocean Sensing and Institute of Quantum Sensing, School of Physics, Zhejiang University, Hangzhou, 310027, China}
\maketitle
\section{Introduction}
\label{sec1}
Owing to their high sensitivity and stability~\cite{Berrada2013,Hardman2016,Fu2019,Heine2020}, 
the Kasevich-Chu (KC) atom gravimeter based on matter-wave interference~\cite{Kasevich1992,Weiss1994}, offers significant scientific value and engineering advantages, for example, applications such as geophysics, mineral exploration, and inertial navigation~\cite{Bidel2018,Zhang2021,Wang2022,Li2023,Chen2025}. 
On the mobile platforms subject to strong decoherence, 
uncorrelated cold atomic ensembles are typically employed,
as a trade-off, to simultaneously achieve high sensitivity and large dynamical bandwidth. According to Fisher information theory, the sensitivity scales as $\delta g\propto 1/(\sqrt{N} k_0 (T/2)^2)$, corresponding to the standard quantum limit. 
Generally, strategies to systematically enhance the sensitivity include improving phase coherence (e.g., suppressing vibration noise, increasing effective recoil momentum $k_0$ and interrogation time $T$)~\cite{Peters2001,Braun2018,Mazzoni2015,Clade2009,McDonald2013}, and enlarging the atom flux $N$~\cite{Colangelo2017,Linnemann2016}. 
However, progresses along these directions are currently saturated by various experimental limitations.

For stationary platforms, a more promising approach is to develop ultracold atom interferometers based on entangled atomic ensembles. 
The proper utilization of many-body entanglement gives new sensing capabilities unachievable with state-of-the-art devices~\cite{Szigeti2021}. 
In particular, many-body entangled states can enable sensitivities beyond the standard quantum limit, approaching the Heisenberg limit $\delta g\propto 1/N$, thereby promising  high-sensitivity atom gravimetry with fewer particles~\cite{Schkolnik2015}.

At present, extensive efforts have been devoted to generate squeezing via nonlinear interactions, including spin squeezing~\cite{Ma2011,SchleierSmith2010} and squeezing of well-separated momentum modes~\cite{Anders2021,Salvi2018}. 
In general, one engineers one-axis or two-axis twisting Hamiltonians~\cite{Ma2011}—using optical cavities~\cite{Greve2022}, spin-exchange interactions in spinor Bose–Einstein condensates~\cite{Hardman2016,Linnemann2016,Kruse2016}, or intercomponent interactions in two-component condensates~\cite{Muntinga2013,Haine2014,Szigeti2020}—to produce spin-squeezed states and thereby achieve appreciable sensitivity enhancement. 
However, many-body entangled states are highly susceptible to noise, making their preparation and detection technically demanding and costly, which hinders their practical deployment. More importantly, compared with high-flux cold-atom interferometers, the sensitivity gain achieved with many-body entangled states has so far remained limited[Ref].

However, most existing studies based on entanglement resources have predominantly focused on spin squeezing, while ignoring a key feature of KC atom interferometers: unlike in their optical counterpart~\cite{Pezze2008,Kong2013}, the two interferometric arms in a KC atom interferometer are formed by a hybridization of internal and external atomic degrees of freedom, analogous to spin–orbit coupling in condensed matter~\cite{Bihlmayer2022}. Consequently, beyond exploiting the spin (intrinsic) degree of freedom to generate squeezing or many-body entanglement, this hybrid structure suggests that squeezing of the external (motional) degrees of freedom may also have a significant impact on the interferometer’s performance.
Indeed, as pointed out in Ref.~\cite{Kritsotakis2018}, in addition to the conventional strategy of enhancing phase coherence, the sensitivity of a KC atom interferometer can also be improved by increasing quantum fluctuations of momentum or position. 
A promising candidate to engineer such fluctuations is motional squeezing state (MSS)~\cite{Morinaga1999}. 
Such state has demonstrated broad impact across a variety of platforms, including Rydberg atoms~\cite{Kaufman2012,Bharti2024}, trapped ions~\cite{Sutherland2021,Shinjo2021,Metzner2024}, and phononic systems, where they enable enhanced control over quantum fluctuations and correlations.
Particularly, in quantum optics, MSS is also referred to as SU(1,1) coherent state~\cite{Caves1981,Pezze2008,Sanders2012}. Its incorporation can enable parameter estimation beyond the standard-quantum limit, as already applied in large-scale interferometric setups such as LIGO~\cite{Caves1981}. 

To date, 
rare attentions have been paid to the role of MSS in atom interferometers~\cite{Linnemann2017,Huang2019},  particularly in KC atom gravimeters. 
In fact, there already exists several well-established experimental techniques for engineering squeezing or stretching of atom position and momentum. For example, frequency-chirped modulation schemes can be employed to generate single-mode or two-mode MSSs~\cite{Sutherland2021,Xin2021,Leong2023,Metzner2024}, and even truncated MSSs can be spontaneously induced with the assistance of Bose–Einstein condensates~\cite{Shukla2025}. 
Moreover, in case of engineering strong spin-squeezing in two-component Bose–Einstein condensates, the free expansion dynamics not only lead to strong spin squeezing, but can also convert interaction energy into kinetic energy~\cite{Castin1996,Stenger1999,Szigeti2020}, resulting in substantial momentum broadening. Both effects can significantly improve the sensitivity of atom interferometers based on two-component Bose–Einstein condensates. 

To systematically elucidate the impact of quantum fluctuations in the external degrees of freedom on the sensitivity of KC atom gravimeters, we utilize SU(2) Lie group formalism~\cite{Yu2025} to construct the unitary evolution operator for arbitrary momentum or position fluctuations, and derive a generic computational framework for both the quantum Fisher information (QFI) and classical Fisher information (CFI).  Using this framework, we quantitatively analyze the maximum achievable sensitivity and optimal control parameters of arbitrary single-mode MSSs under two commonly-employed measurement schemes, population measurement and joint population–position measurement~\cite{Dickerson2013,Sugarbaker2013,Hardman2016,Kritsotakis2018,Wang2023}. 
In particular, we investigate the strong Doppler effects induced by large momentum broadening. We find that MSSs with large position fluctuations or extended coherence length can provide, at most, a fourfold sensitivity enhancement over the semiclassical limit under population measurement. In contrast, the joint population–position measurement yields substantially higher sensitivity gains over a broader range of squeezing parameters. 
Furthermore, although Doppler effects inevitably suppress the sensitivity enhancement provided by motional squeezing, our quantitative analysis shows that their impact on the gains associated with large quantum fluctuations is bounded. As a result, MSSs can still deliver significant sensitivity improvements, especially under joint measurement of population–position. Finally, we briefly discuss pulse imperfections arising from Doppler effects. 
It is worthy noting here that the observed sensitivity enhancement thorough this work is always limited to the quantum standard limit because there is no many-body entanglement. 

The rest of this manuscript is arranged as follow. 
In the second section, we derive a generic computational framework of quantum and classical Fisher information of two commonly-deployed measurement protocols. 
In the third section, we use this framework to analyze the sensitivity of KC atom interferometer with MSSs as an input, and demonstrate in detail that the sensitivity enhancement from quantum quadrature fluctuations and the corresponding Doppler suppression from large momentum broadening. 
Finally, we make a conclusion and outlook for this study. 

\section{Computational Framework of Quantum and Classical Fisher information}
\label{sec2}
The relative scarcity of applications of MSSs in atom interferometry can be partly attributed to the non-negligible Doppler effects associated with most such states, which prevent the light pulses from realizing ideal $\pi/2$ or $\pi$ operations~\cite{Saywell2018,Wilkason2022}.  
In fact, the significant impact of Doppler effects from finite-temperature cold atoms on conventional KC atom gravimeters is well known, yet the context of this work is completely different. 
To explicitly account for Doppler effects, we formulate the unitary evolution operators under both “Ideal” and “Doppler” scenarios, and establish corresponding unified frameworks for evaluating the QFI and CFI. The “Ideal” scenario assumes perfect light pulses, whereas the “Doppler” scenario fully incorporates pulse imperfections induced by Doppler effects. For the latter, we employ the SU(2) Lie group formalism to derive a generic unitary evolution operator ${\hat U}_{\rm KC}$. 
It should be noted that we do not address the problem of optimal measurements associated with MSSs here; the rationale will be discussed later.
Without loss of generality, the input state of a KC atom interferometer can be written as a product of internal state $|a\rangle$ and external one $|\Psi_0\rangle$, that is, $|a\rangle |\Psi_0\rangle$. 
The final state is then given by the unitary evolution $|\Psi(t)\rangle={\hat U}_{\rm KC}|a\rangle|\Psi_0\rangle$,  
where the gravitational slope $g$ is encoded into the output state via the Hamiltonian. Since in our case the transverse motion does not affect the sensitivity, we restrict our analysis to the direction alongside the linear  gravitational potential.

\subsection{Quantum Fisher Information}
The lowest bound of sensitivity of atom interferometry is determined by the QFI, also termed as quantum Cramer-Rao bound (QCRB) in the literature~\cite{Braunstein1994}. 
For $N$ uncorrelated particles, this is $\delta g^2\geq 1/(NF_Q)$, $F_Q=4(\langle \partial_g\Psi(t)|\partial_g\Psi(t)\rangle-|\langle\Psi(t)|\partial_g\Psi(t)\rangle|^2)$ is denoted as the single-particle QFI. 
To clearly show the Doppler effects, we divide the discussions into two scenarios, "Ideal" (notated as $\mathcal{S}={\mathcal I}$) and "Doppler" ($\mathcal{S}={\mathcal D}$) scenario, as stated in the last paragraph. 
In both scenarios, the evolution operator can be written down as follow,
\begin{eqnarray}
	{\hat U}_{\rm KC}^{\mathcal S}&=&{\hat V}_3^{\mathcal S} {\hat U}_g(T_2) {\hat V}_2^{\mathcal S} {\hat U}_g(T_1) {\hat V}_1^{\mathcal S},\nonumber\\
	{\hat V}_j^{\mathcal S}&=&{\hat U}_0 {\tilde U}_j^{\mathcal S} {\hat U}_0^\dagger,
	\label{UKC}
\end{eqnarray}
here both $\hat U_0=|a\rangle\langle a|\exp(-ik_0{\hat z}/2)+|b\rangle\langle b|\exp(ik_0{\hat z}/2)$ and ${\hat U}_0^\dagger$ together describes the momentum transfer ($\hbar k_0$) from the laser pulse, while ${\hat U}_g(T_{1,2})$ includes the free falling and expansion in the gravitational potential, 
\begin{eqnarray}
	{\hat U}_g(T_{1,2})&=&{\hat U}_p(T_{1,2})\exp(ig{\hat G}(T_{1,2})),\\
	{\hat G}(T)&=&\frac{mT}{\hbar} {\hat z}+\frac{T^2}{2\hbar} {\hat p},
\end{eqnarray}
here ${\hat U}_p(T_{1,2})=\exp(-i\frac{T_{1,2}}{\hbar}\frac{{\hat p}^2}{2m})$. $T_1$ and $T_2$ denotes as the first and second interrogation time, respectively (always satisfying $T_1+T_2=T$). 
For the conventional light-atom interaction, 
the shift from the momentum broadening of input state is negligible, $\Delta p k_0/m\ll \Omega$, 
that is the "Ideal" scenario. 
Then ${\tilde U}_j^{\mathcal{I}}$ can be written down as follow,
\begin{eqnarray}
	\label{eqn:UjI}
	{\tilde U}_{j}^{\mathcal I}&=&\cos(\Omega \tau_j/2){\hat I}\\ 	
	&&-i\sin(\Omega \tau_j/2)
	\cdot\big(e^{i\phi_j}|a\rangle\langle b|+e^{-i\phi_j}|b\rangle\langle a|\big),\nonumber
\end{eqnarray}
$\Omega\tau_j=\pi/2$ and $\pi$ ($\Omega$ is the Rabi frequency) corresponds to the $\pi/2$ and $\pi$ pulse, respectively. And $\phi_j$ is the phase of $j$th laser pulse, $|a\rangle$ and $|b\rangle$ are denoted as the two internal states. 
However, this evolution operator (Eq.~\ref{eqn:UjI}) becomes inaccurate for input state with large momentum fluctuations, i.e., $\Delta p k_0/m\gtrsim \Omega$ where strong Doppler effects are non-negligible. 
According to our previous framework based on SU(2) Lie group theory~\cite{Yu2025}, the unitary operator ${\tilde U}^{\mathcal{D}}_j$ describing the imperfect interaction between laser pulses and two-level atom under "Doppler" scenario can be expressed as
\begin{eqnarray}
	\label{eqn:UjD}
	{\tilde U}^{\mathcal{D}}_j({\hat p})={\hat U}^{\mathcal{D}}_j({\hat p}-mgt),
\end{eqnarray}
with 
\begin{eqnarray}
	{\hat U}^{\mathcal{D}}_j({\hat p})
	&=&\big({\hat A}_j+i{\hat B}_j\big)|a\rangle\langle a|+\big({\hat A}_j-i{\hat B}_j\big)|b\rangle\langle b|\\ 
	&&-i\hat C_j\big(\exp(i\phi_j)|a\rangle\langle b|+\exp(-i\phi_j)|b\rangle\langle a|\big),\nonumber
\end{eqnarray}
$\hat A_j=\cos(\hat\Delta \tau_j/2)$, $\hat B_j=\hat b_0/{\hat \Delta}\sin(\hat\Delta \tau_j/2)$ and $\hat C_j=\Omega/{\hat \Delta}\sin(\hat\Delta \tau_j/2)$ with $\hat\Delta=\sqrt{{\hat b}_0^2+\Omega^2}$, $\hat b_0= k_0{\hat p}/m+\delta_0$ with the initial detuning $\delta_0$. Here it is worthy noting that Eq.~\ref{eqn:UjD} is valid for the case that the chirping rate of laser pulses is $\delta(t)=\delta_0-k_0gt$, which exactly compensates for the Doppler shift arising from the uniform acceleration of atoms in a linear potential~\cite{Cheng2015}. It is straightforward to see that when the Rabi frequency $\Omega$ is much larger than all other typical frequency scales, i.e.,  $\Delta p k_0/m$, Eq.~(\ref{eqn:UjD}) reduces to Eq.~(\ref{eqn:UjI}).

In the "Ideal" scenario,  the single-particle QFI $F_{\rm Q}=F_{\rm Q}^{\mathcal I}$ can be recast as follow~\cite{Kritsotakis2018},
\begin{eqnarray}
	\label{eqn:QFI-Ideal-0}
	F_{\rm Q}^{\mathcal I}=k_0^2\big(\frac{T^2}{2}-T_2^2\big)^2+4{\rm Var}\big({\hat G(T)}\big).	
\end{eqnarray}
As a result, both items in the right-hand-side of Eq.~\ref{eqn:QFI-Ideal-0} can enhance the sensitivity of KC atom interferometry. However, the first item originates from the phase coherence of the two interference arms, while the second term stems from the external quantum fluctuations of the input state, ${\rm Var}(\cdot)$ denotes as the variance. 
For the "Doppler" scenario, the calculation of the QFI is lengthy and tedious. Here we only display the final results (For the details, please go to the Appendix~\ref{appendix:QFI-Doppler}), 
\begin{eqnarray}
	\label{eqn:QFI-Doppler-0}
	F_{\rm Q}^{\mathcal D}
	=k_0^2\big(\frac{T^2}{2}-T_2^2\big)^2+4\langle\Psi_0| \hat G^{\prime\prime}|\Psi_0\rangle+4{\rm Var}({\hat G}^\prime).
\end{eqnarray}
$\hat G^\prime={\hat d}_{0,0}+\frac{mT}{\hbar}{\hat z}$, 
where both ${\hat d}_{0,0}$ and ${\hat G}^{\prime\prime}$ are operators defined in momentum space, also listed in the Appendix~\ref{appendix:QFI-Doppler}. 
When $\Omega\rightarrow \infty$, $\hat G^{\prime\prime}\rightarrow 0$ and ${\rm Var}({\hat G}^\prime)\rightarrow{\rm Var}\big({\hat G(T)}\big)$, these coincide with the "Ideal" scenario, 
that is, $\lim_{\Omega\rightarrow\infty}F_{\rm Q}^{\mathcal{D}}=F_{\rm Q}^{\mathcal{I}}$. 
When neglecting the quantum fluctuations of the input state and setting \textit{interrogation timing-symmetry} $T_1=T_2=T/2$, both Eq.~\ref{eqn:QFI-Ideal-0} and  Eq.~\ref{eqn:QFI-Doppler-0} consistently give the semi-classical limit (SCL), $F_{\rm SCL}=k_0^2(T/2)^4$. 
Comparing the QFIs between "Doppler" and "Ideal" scenario, 
two distinct features in the "Doppler" case are summarized here and the details are also left in the Appendix~\ref{appendix:QFI-Doppler}. 
Firstly, the external quantum fluctuations (${\hat G}^\prime$ and ${\hat G}^{\prime\prime}$) in the "Doppler" scenario depend both on the interrogation time $T_1$ and $T_2$, rather than the total time $T=T_1+T_2$. 
Secondly, the QFI (Eq.~\ref{eqn:QFI-Doppler-0}) under "Doppler" scenario is weakly affected by the gradient of gravitational potential $g$ and the phase difference of pulsed lasers through a phase $\Theta$ (defined in momentum space), 
\begin{eqnarray}
	\hat\Theta=\phi_{1}-\phi_2+\frac{k_0T_1}{m}(\hat p+\frac{\hbar k_0}{2})+\frac{gk_0T_1^2}{2}.
	\label{eqn:Theta}
\end{eqnarray}

While increasing the momentum fluctuation or broadening may potentially enhance the sensitivity (Eq.~\ref{eqn:QFI-Ideal-0} and Eq.~\ref{eqn:QFI-Doppler-0}), it also causes a fraction of atoms to deviate from the main interference paths of the KC atom interferometer. As a result, the number of atoms that actually participate in the KC interferometer is reduced, ultimately weakening the sensitivity gain brought by the increased momentum fluctuation. To characterize the impact of large momentum broadening, we introduce the fidelity between two scenarios, defined as $\mathcal{F}=|\langle\Psi_0|\langle a|{\hat U}_{\rm KC}^{\mathcal{I}\dagger}{\hat U}_{\rm KC}^{\mathcal{D}}|a\rangle|\Psi_0\rangle|$, which clearly reflects the imperfection of light pulses induced by the Doppler broadening, 
\begin{eqnarray}
	\label{eqn:fidelity}
	{\mathcal F}&=&|\langle\Psi_0|\langle a|e^{-i\frac{k_0{\hat z}}{2}}{\tilde  U}_{1}^{\mathcal{I}\dagger}e^{-i\hat\beta_1\sigma_z}{\tilde  U}_2^{\mathcal{I}\dagger}e^{-i\hat\beta_2\sigma_z}{\tilde U}_{3}^{\mathcal{I}\dagger}\nonumber\\
	&&{\hat U}_3^{\mathcal{D}}e^{i\hat\beta_2\sigma_z}{\hat U}_2^{\mathcal{D}}e^{i\hat\beta_1\sigma_z}{\hat U}_1^{\mathcal{D}}e^{i\frac{k_0{\hat z}}{2}}|a\rangle|\Psi_0\rangle|, 
\end{eqnarray}
here $\hat\beta_1=gk_0T_1^2/4+k_0T_1{\hat p}/(2m)$ and $\hat\beta_2=gk_0(T_1^2+2T_1T_2)/4+k_0T_2{\hat p}/(2m)$.

\subsection{Classical Fisher Information}
Although the QFI quantifies the maximum amount of information that can be extracted from a measurement, achieving the highest sensitivity in practice typically requires an optimal measurement scheme, i.e., one that saturates the QCRB. 
In the “Ideal" scenario, it has been pointed out in the literature that applying a harmonic potential before measurement to implement an appropriate rotation in phase space enables an optimal measurement. In this way, the corresponding CFI can reach the QFI, thus saturating the QCRB~\cite{Kritsotakis2018}. 
However, this scheme is difficult to generalize to the 
"Doppler" scenario, i.e., the input MSS with large momentum broadening. 
Three main challenges arise here: (i), Due to the finite momentum transfer of pulsed lasers, a spin-dependent momentum kick is required to eliminate the velocity difference between different internal states; this becomes difficult to implement in the presence of large momentum broadening. 
(ii), Because of free expansion, the spatial extent of the atomic cloud can becomes too large to apply a suitable harmonic trapping potential prior to measurement.
(iii), The Doppler effect can, to some extent, compromise the universality of the phase-space rotation operation~\cite{Yu2025}.
Taking these facts into account, we do not address the problem of optimal measurement for MSS in this work. Instead, we focus on two types of experimentally feasible measurements: population measurement (denoted as $\{\hat S_z\}$) and joint measurement of population and position ($\{\hat z, \hat S_z\}$). We then explicitly compute the CFI for arbitrary momentum or position fluctuations. 
The details are discussed in below. 

\subsubsection{Population Measurement and the Classical Fisher Information}
The population measurement is the most common employed scheme in practical applications, which can be quantified by the corresponding CFI, $F_{\rm C}^{\mathcal{I},\mathcal{D}}(\{\hat S_z\})=\sum_{s=a}^b(\partial_g P_s)^2/P_s$, here the information on $g$ is encoded in the probability $P_s$ of $|a\rangle$ and $|b\rangle$.

Firstly in the "Ideal" scenario, the probability can be recast analytically $P_{a/b}=(1\pm {\mathcal{C}\cos(\varphi)})/2$ with $\varphi=\Phi-\phi_f-\phi_g$, 
here $\Phi=\phi_1-2\phi_2+\phi_3$, 
$\phi_f=E_{\rm r}(T_2-T_1)/\hbar$ and $\phi_g=\frac{gk_0}{2}(T^2-2T_1^2)$ quantifies the total phase difference, the phase shift from the linear potential and the one from momentum kicks of pulsed lasers, respectively. Here $E_{\rm r}=\hbar^2 k_0^2/(2m)$ denotes the recoil energy induced by laser pulses. 
Therefore, the CFI in the "Ideal" scenario, 
\begin{eqnarray}
	\label{eqn:CFI-Ideal-Sz}
	F_{\rm C}^{\mathcal{I}}(\{\hat S_z\})=\frac{k_0^2(T^2-2T_1^2)^2}{4}{\mathcal C}^2, 
\end{eqnarray}
in which the optimal phase difference  $\Phi^{\mathcal{I}}(\{\hat S_z\})=\phi_f+\phi_g+\pi/2$ is set. 
And the contrast of interference pattern,
\begin{eqnarray}
	\label{eqn:contrast}
	{\mathcal C}=\langle \Psi_0|\exp(i\frac{k_0}{m}(T_2-T_1){\hat p})|\Psi_0\rangle,
\end{eqnarray}
which is the wavefunction overlap from the distinct interference arms. 

Secondly in the "Doppler" scenario, even though the expression of probability $P_{a/b}$ is lengthy, 
the CFI can be expressed explicitly by introducing the interrogation timing-asymmetry $r\equiv T_1/T$ ($r=1/2$ corresponds to symmetry point) as follow, 
\begin{eqnarray}
	\label{eqn:CFI-Doppler-Sz}
	F_{\rm C}^{\mathcal{D}}(\{\hat S_z\})=
	\frac{k_0^2T^4}{4}\sum_s \frac{\left(M_s+r^2N_s\right)^2}{P_s},
\end{eqnarray}
here
\begin{eqnarray}
	P_s&=&\langle\Psi_0|{\hat K}_{sa}^\dagger {\hat K}_{sa}|\Psi_0\rangle,\nonumber\\
	M_s&=&\frac{i}{2}\langle\Psi_0|{\hat K}_{sa}^\dagger {\hat K}_{sa}^\prime-{\hat K}_{sa}^{\prime\dagger} {\hat K}_{sa}|\Psi_0\rangle,\nonumber\\
	N_s&=&\frac{i}{2}\langle\Psi_0|{\hat K}_{sa}^\dagger {\hat K}_{sa}^{\prime\prime}-{\hat K}_{sa}^{\prime\prime\dagger} {\hat K}_{sa}|\Psi_0\rangle.
\end{eqnarray}
and
\begin{align}
	{\hat K}_{sa}=e^{-i\frac{k_0{\hat z}}{2}}\langle s|{\hat U}_3^{\mathcal D}e^{i\hat\beta_2\sigma_z}{\hat U}_2^{\mathcal D}e^{i\hat\beta_1\sigma_z}{\hat U}_1^{\mathcal D}|a\rangle e^{i\frac{k_0{\hat z}}{2}},\nonumber\\
	{\hat K}_{sa}^\prime = e^{-i\frac{k_0{\hat z}}{2}} \langle s| {\hat U}_3^{\mathcal{D}}e^{i\hat\beta_2\sigma_z}\sigma_z{\hat U}_2^{\mathcal{D}}e^{i\hat\beta_1\sigma_z}{\hat U}_1^{\mathcal{D}}|a\rangle e^{i\frac{k_0{\hat z}}{2}},\nonumber\\
	{\hat K}_{sa}^{\prime\prime}=e^{-i\frac{k_0{\hat z}}{2}} \langle s| {\hat U}_3^{\mathcal{D}}e^{i\hat\beta_2\sigma_z}[{\hat U}_2^{\mathcal{D}},\sigma_z]e^{i\hat\beta_1\sigma_z}{\hat U}_1^{\mathcal{D}}|a\rangle e^{i\frac{k_0{\hat z}}{2}},\nonumber\\
\end{align}
$\sigma_z=[1,0;0,-1]$ is the third Pauli matrix. A similar equivalence can be proven out as $\Omega\rightarrow\infty$, $\lim_{\Omega\rightarrow\infty} F^{\mathcal{D}}_{\rm C}(\{{\hat S}_z\})= F_{\rm C}^{\mathcal I}(\{{\hat S}_z\})$. 

\subsubsection{Joint measurement of Population-position and the Classical Fisher Information}
Compared with the population measurement, 
joint measurement of population-position enables much better performance and more possibilities for a KC atom interferometer, due to the additional information of spatial distribution.
Actually, Joint measurement of Population-position is broadly investigated and applied in laboratory and mobile platforms, i.e., Point-source interferometry or Phase-shear interferometry~\cite{Sugarbaker2013,Dickerson2013,Chen2019,Avinadav2020}. 
In the later case, the phase is mapped into the probability distribution to realize multi-parameter estimation of rotation and the gradient of the linear potential, by introducing phase shear and interrogation timing-asymmetry along the transverse and longitudinal direction, respectively. 
The single-particle CFI of joint measurement of population and position is estimated according to the probability distribution $P_s(z)$ with $P_{s}(z)=|L_{s}(z)|^2$, 
\begin{eqnarray}
	L_s(z)=\langle z-z_0|{\hat K}_{sa}{\hat U}_p(T)|\Psi_0\rangle.
\end{eqnarray}
here $z_0=\frac{gT^2}{2}+\frac{\hbar k_0T}{2m}$ represents  the mean position at the output from free falling and drift of momentum kicks. 
The CFI by definition can be written down as $F^{{\mathcal I},{\mathcal D}}_{\rm C}(\{{\hat z}, {\hat S}_z\})=\sum_{s=a}^b\int dz \frac{\left(\partial_g P_s(z)\right)^2}{P_s(z)}$, and simplified as follow, 
\begin{eqnarray}
	\label{eqn:CFI-zSz}
	F^{\mathcal I, \mathcal D}_{\rm C}(\{{\hat z}, {\hat S}_z\})=\sum_{s=a}^b \int dz \frac{k_0^2T^4}{4L_s^*L_s}
	\left({\rm Im}(L_sH_s^*)\right)^2,
\end{eqnarray}
here
\begin{align}
	H_s(z)=\langle z|\Big({\hat K}_{sa}^\prime-\frac{2{\hat p}}{\hbar k_0}{\hat K}_{sa}+r^2{\hat K}_{sa}^{\prime\prime}\Big){\hat U}_p(T)|\Psi_0\rangle.
\end{align}
In the "Ideal" scenario, it can be proven that $\lim_{\Omega\rightarrow\infty} F^{\mathcal{D}}_{\rm C}(\{{\hat z}, {\hat S}_z\})= F_{\rm C}^{\mathcal I}(\{{\hat z}, {\hat S}_z\})$. 

We note here that the difference of both scenarios on the phases of laser pulse. 
In the "Ideal" scenario, both the two CFIs (Eq.~\ref{eqn:CFI-Ideal-Sz} and Eq.~\ref{eqn:CFI-zSz}) depend on 
the total phase difference $\Phi=\phi_1-2\phi_2+\phi_3$, while in the "Doppler" scenario, both the two CFIs (Eq.~\ref{eqn:CFI-Doppler-Sz}, Eq.~\ref{eqn:CFI-zSz}) depend on $\psi_{12}=\phi_1-\phi_2$ and $\psi_{23}=\phi_2-\phi_3$ individually. 

\section{Enhanced sensitivity and Doppler supression of Motional squeezing states}
\label{sec3}
The discussions on MSS or SU(1,1) coherent state have a long history, and their applications are crucial to many advanced setups aiming to measure extremely weak signals. 
In recent decades, the discussions on its analogy in matter-wave field attract more attentions. 
Assuming that the atoms are prepared in a one-dimensional harmonic trapping potential, 
the input external state $|\Psi_0\rangle$ is set to be a MSS, which can be 
expressed by the squeezing operator acting on the vacuum state of a harmonic trapping potential $|0\rangle$, 
$|\Psi_0\rangle=\exp(\xi {\hat a}^{\dagger2}/2-\xi^*{\hat a}^2/2)|0\rangle$~\cite{Sanders2012,Huang2019}, $\xi=\beta\exp{(i\theta)}$, here ${\hat a}$ and ${\hat a}^\dagger$ is the creation and annihilation operator of a harmonic potential, respectively. The MSS in momentum space can be written down as follow, 
\begin{eqnarray}
	\label{eqn:Psi0}
	\langle p|\Psi_0\rangle=\frac{(1-|\alpha|^2)^{1/4}}{\sqrt{\sigma_p\sqrt{\pi}}\sqrt{1-\alpha}}e^{\left(-\frac{1+\alpha}{2(1-\alpha)}\frac{p^2}{\sigma_p^2}\right)}.
\end{eqnarray}
$\alpha=\tanh(\beta)e^{i\theta}$ is the parameter defined in the Poincar$\acute{\rm e}$ disk, $|\alpha|\leq 1$. 
As we will see in the remained, this nonlinear phase on the momentum of MSS (Eq.~\ref{eqn:Psi0}) takes significant roles. 
The squeezing strength and direction in phase space is controlled by $(\beta,\theta)$. $\sigma_p=\sqrt{m\hbar\omega}$ and $\omega$ is the momentum broadening and frequency of the harmonic trapping, respectively. The momentum and position broadening, $\Delta p$ and $\Delta z$, can be recast as follow, 
\begin{eqnarray}
	\Delta p&=&\frac{\sigma_p}{\sqrt{2}}\frac{|1-\alpha|}{\sqrt{1-|\alpha|^2}},\\
	\Delta z&=&\frac{\hbar}{\sqrt{2}\sigma_p}\frac{|1+\alpha|}{\sqrt{1-|\alpha|^2}},
\end{eqnarray}
Correspondingly, the Heisenberg uncertainty then is $\Delta p(\alpha)\Delta z(\alpha)=\hbar|1-\alpha^2|/(2(1-|\alpha|^2))$. 
Apparently, each quadrature fluctuation is symmetric about  the real axis of $\alpha$ (Im$(\alpha)=0$), 
where the Heisenberg  uncertainty achieves the minimum. 
Around $\alpha=1$, the position (momentum) fluctuation is extremely magnified (suppressed), as a comparison, complete opposite behaviors are expected around $\alpha=-1$. 
Thereby, the Doppler scenario is indeed needed as close to $\alpha=-1$ while the Ideal scenario is enough as close to $\alpha=1$. 
The tunable squeezing of MSS enable us to investigate the quantum external fluctuations of the input state.  

Before discussing the CFIs of two chosen measurement protocols, it would be heuristic to discuss the corresponding QFIs in the two scenarios. 
In the "Ideal" scenario, the term from quantum fluctuation of input MSS can be calculated analytically,
\begin{align}
	\frac{8{\rm Var}(\hat G)}{k_0^2T^4}
	=\big(\frac{\sigma_p}{\hbar k_0}\sin(\theta/2)+\frac{\sigma_T}{\sigma_p}\cos(\theta/2)\big)^2e^{2\beta}\nonumber\\
	+\big(\frac{\sigma_p}{\hbar k_0}\cos(\theta/2)-\frac{\sigma_T}{\sigma_p}\sin(\theta/2)\big)^2e^{-2\beta}.
\end{align}
$\sigma_T=2m/(k_0T)$, $\hbar/\sigma_T$ quantifies the position uncertainty of output atoms induced by momentum kicks in the frame of free-falling. 
It is obvious that QFI can be exponentially enhanced along a proper direction in the Poincare Disk. 
Actually, in case of a fixed $\beta$, $F_Q^{\mathcal{I}}$ (Eq.~\ref{eqn:QFI-Ideal-0}) achieve the maximum and minimum when $\theta=(\theta_c,\theta_c+\pi)$, 
\begin{eqnarray}
	{\rm Var}(\hat G(T))\Big|^{\rm Max}_{\rm Min}=\frac{k_0^2T^4}{2}e^{\pm 2\beta}\left(1+\left(\frac{2}{\omega T}\right)^2\right),
\end{eqnarray}
respectively. Here 
\begin{eqnarray}
	\theta_c=2\arctan\left(\frac{\omega T}{2}\right),
\end{eqnarray}
Generally, ${\rm Var}(\hat G(T))$ (also $F_{\rm Q}^{\mathcal{I}}$ Eq.~\ref{eqn:QFI-Ideal-0}) only exhibits \textit{symmetry} about Re$(\alpha)=0$ when $\theta_c=\pi/2$ for $\omega T=2$. 
While in the "Doppler" scenario, the expression of $F_{\rm Q}^{\mathcal D}$ (Eq.~\ref{eqn:QFI-Doppler-0}) does not admit further analytical simplification and therefore requires numerical evaluation, even for a Gaussian input state (Eq.~\ref{eqn:Psi0}). 

As both the encoding scheme (input state) and decoding scheme (measurement protocol) are fixed in our KC atom interferometer, the maximum sensitivity for each measurement protocol can be achieved by optimizing the control parameters, including the interrogation-time ratio $r\equiv T_1/T$ and the laser phases $\phi_j$~\cite{Oh2019,Ataman2022,ZhouSisi2023,Belliardo2024,Ullah2025}. 
Specifically, using the single-particle QFI/CFI framework developed in Sec.~\ref{sec2}, we determine the global maximum of the corresponding CFI for each scenario and measurement protocol, analytically or numerically, to identify the optimal parameters (e.g., $r$ and the phase difference $\Phi$ in the "Ideal" scenario), and subsequently evaluate the QFI at these optimized settings. 
In the following two subsections, by comparing the two measurement protocols in both the "Ideal" and "Doppler" scenarios, we demonstrate a significant sensitivity enhancement enabled by the quantum fluctuations of the input MSS, as well as the robustness of this enhancement against Doppler suppression. 

\subsection{Population measurement}
\begin{figure}
	\centering
	\includegraphics[width=1.0\linewidth]{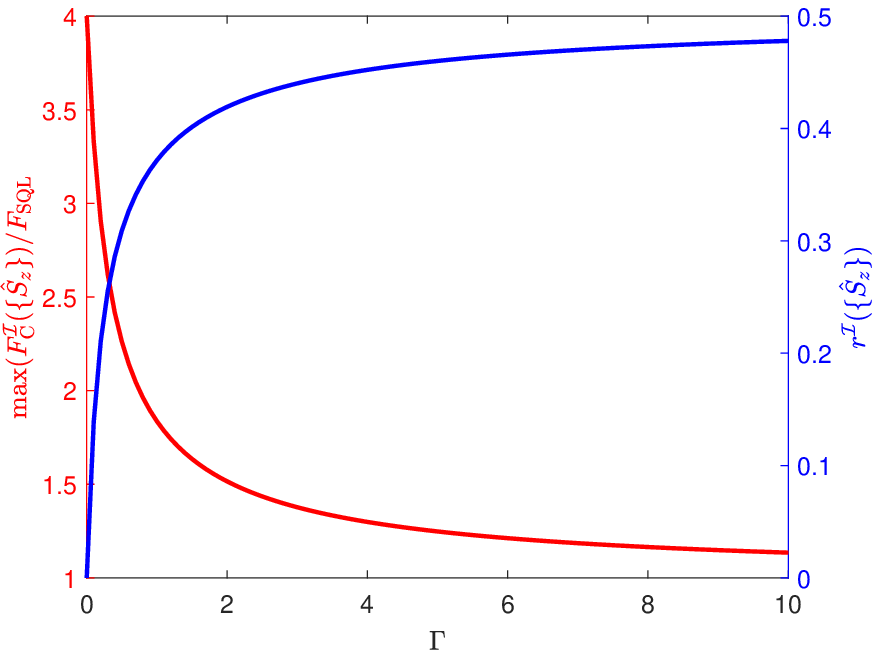}
	\caption{In the "Ideal" scenario, the global maximum CFI of population measurement ($\{{\hat S}_z\}$) and the corresponding optimal ratio $r^{\mathcal{I}}(\{{\hat S}_z\})$ versus $\Gamma$.}
	\label{fig1}
\end{figure}
In a KC atom interferometer with population measurement, 
as seen in Eq.~\ref{eqn:CFI-Ideal-Sz}, the maximum sensitivity are determined by two competing mechanisms, one is phase coherence of the two interference arms which is quantified by their area difference $\Delta S=\frac{\hbar k_0}{m}|\frac{T^2}{2}-T_2^2|$~\cite{Storey1994}, the other is the contrast of interference pattern which is exponentially affected by the timing-asymmetry $T_1\neq T_2$ or equivalently $r\neq 1/2$. The former favors the optimal ratio around $r=0$ or $r=1$, while the later tends to $r\approx 1/2$.  Apparently, the conventional ratio $r=1/2$ can not be applied to the MSS case as usual. 

In "Ideal" scenario, the CFI of population measurement is determined by the constrast $\mathcal{C}$ (Eq.~\ref{eqn:contrast}), which is obtained analytically ${\mathcal{C}}=\exp(-(1-2r)^2\Gamma)$ with
\begin{eqnarray}
	\Gamma=\frac{\sigma_p^2}{\sigma_T^2}\left(\big(e^{\beta}\sin(\theta/2)\big)^2+\big(e^{-\beta}\cos(\theta/2)\big)^2\right).
\end{eqnarray}
Generally, it can be proven that $F_C^{\mathcal{I}}$ (Eq.~\ref{eqn:CFI-Ideal-Sz}) achieves the extremum  of $F_C^{\mathcal{I}}(\{\hat S_z\})$ 
when 
\begin{eqnarray}
	\label{eqn:ratio-Ideal-Sz}
	r=2\Gamma(1-2r)(1-2r^2),
\end{eqnarray}
is satisfied. The solution that maximize $F_C^{\mathcal{I}}(\{\hat S_z\})$  is denoted as $r^{\mathcal{I}}(\{\hat S_z\})$. 
Two features are noted here: i), both $\Gamma$ and $F_{\rm C}^{\mathcal I}(\{\hat S_z\})$ (Eq.~\ref{eqn:CFI-Ideal-Sz}) are  
\textit{symmetric} about Im$(\alpha)=0$; ii), the \textit{symmetry} about $r=1/2$ is also broken by phase coherence.

\begin{figure}
	\includegraphics[width=1.0\linewidth]{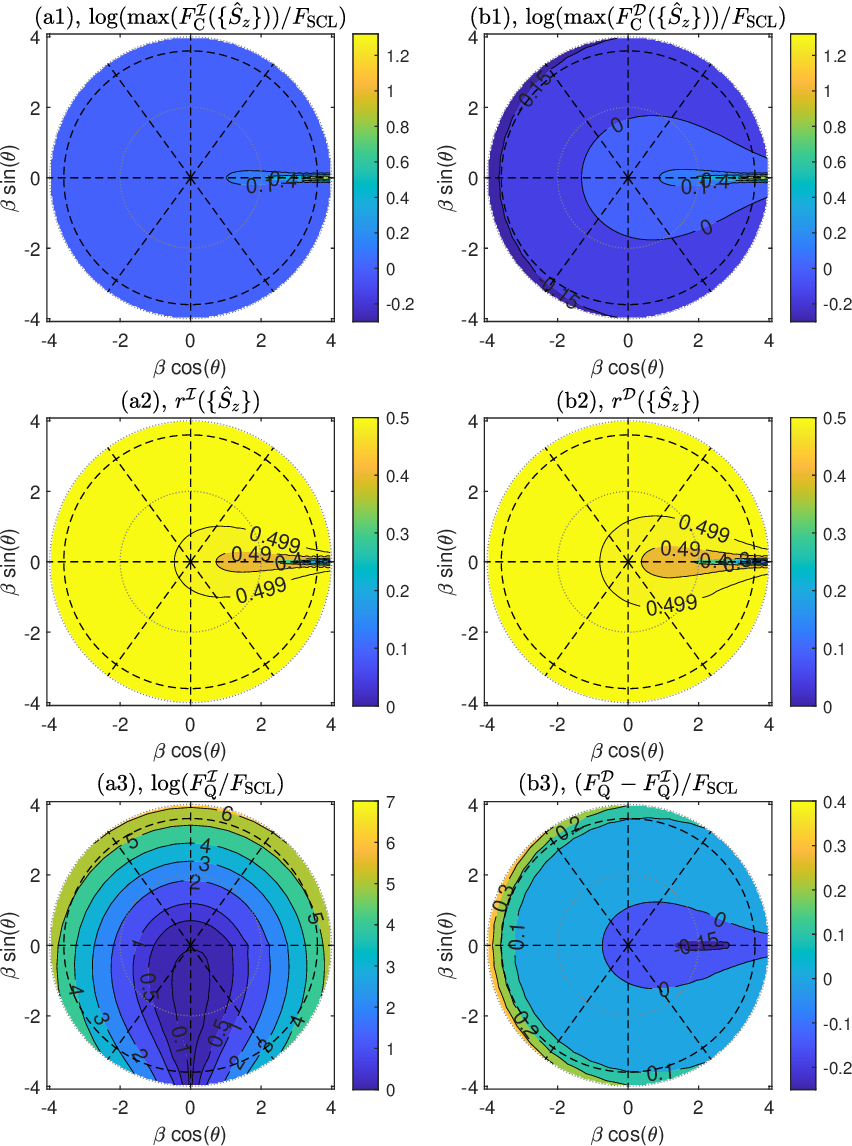}
	\caption{(color online) Evolution of max($F_{\rm C}^{\mathcal{S}}(\{\hat S_z\})$) ((a1,b1), log scale), optimal ratio $r^{\mathcal S}(\{\hat S_z\})$ (a2,b2) and the corresponding QFI $F_{\rm Q}^{\mathcal{S}}$ (a3,b3) versus squeezing parameters $(\beta,\theta)$ for the "Ideal" ($\mathcal{S}=\mathcal{I}$) and "Doppler" ($\mathcal{S}=\mathcal{D}$) scenario. 
	The black dashed radial lines (concentric circle)  correspond to $\theta/\pi=(0,0.3,0.5,0.7,1.0,1.3,1.7,2)$ ($\beta=3.6$). 
	Here $\delta_0=-E_{\rm r}$, $\hbar\Omega/E_{\rm r}=20$, $\omega T=2$, $E_{\rm r}T/\hbar=200$.}
	\label{fig:fig3rd-1}
\end{figure}

By numerically solving Eq.~\ref{eqn:ratio-Ideal-Sz} and finding out the maximum CFI, both evolution of max$(F_{\rm C}^{\mathcal{I}}(\{\hat S_z\}))$ and $r^{\mathcal{I}}(\{\hat S_z\})$ versus $\Gamma$ are present in Fig.~\ref{fig1}. 
In the absence of squeezing ($\beta=0$ whatever $\theta$ is),  $\Gamma=\sigma_p^2/\sigma_T^2$ which usually satisfies $\Gamma\gg 1$, as a result, the optimal ratio $r\approx1/2$ is chosen with $F_{\rm C}^{\mathcal{I}}(\{\hat S_z\})\approx F_{\rm SCL}$ in a conventional setup. 
Actually, this phenomenon persists across the most parameter space ($\beta,\theta$) (see the (blue, yellow) regime in Fig.~\ref{fig:fig3rd-1} (a1,b1)) because the interference contrast consistently dominates the competition. 
However, around the $\theta=0$ sector, the dominant factor is the phase coherence of two interference arms 
because position fluctuation (atom coherence length) can be exponentially enhanced while the momentum fluctuation is tiny as increasing $\beta$~\cite{Parazzoli2012}. 
Although the parameter regime in which max$(F_{\rm C}^{\mathcal{I}}(\{\hat S_z\}))$ becomes large is very narrow, we find that by reducing $\Gamma$ to zero, the max$(F_{\rm C}^{\mathcal{I}}(\{\hat S_z\}))$ (red solid line in Fig.~\ref{fig1} and Fig.~\ref{fig:fig3rd-1}(a1)) can reach a maximum as \textit{four times} the $F_{\rm SCL}$ while $r^{\mathcal{I}}(\{\hat S_z\})\rightarrow 0$, as depicted in Fig.~\ref{fig1}. 

In the "Doppler" scenario where Doppler effects are fully taken into account, 
however, the sensitivity max$(F_{\rm C}^{\mathcal{D}}({\hat S_z}))$ is usually further suppressed, this suppression becomes increasingly pronounced as $(\beta,\theta)$ moving away from $\theta=0$ and approaching to $\theta=\pi$, and eventually falls significantly below the SCL. 
This degradation arises because strong momentum fluctuations prevent a fraction of atoms from effectively participating in the interference process. 
For the optimal ratio, 
we have numerically checked that the difference on optimal ratio $r^{\mathcal{I}}({\hat S_z})$ and $r^{\mathcal{D}}({\hat S_z})$ is minimal (Fig.~\ref{fig:fig3rd-1} (a2,b2)). 
For QFI, we numerically find that the absolute amplitude of QFI difference $\Delta F_{\rm Q}=F_{\rm Q}^{\mathcal{D}}-F_{\rm Q}^{\mathcal{I}}$ is bounded even with strong Doppler effects, e.g. $k_0\Delta p /(m\Omega)\approx 0.4$ when $\beta=4.0$, $\theta=\pi$ and $\sigma_p/(\hbar k_0)=0.1$. 
However, 
around the regimes $\theta=\theta_c=\pi/2$ and $\beta\leq 1$ (the deep blue regime) in Fig.~\ref{fig:fig3rd-1} (a3,b3) where $F_{\rm Q}^{\mathcal I}\approx F_{\rm SCL}$, the relative amplitude of QFI difference is large enough to be considered in. 

\subsection{Joint measurement of population and position}
In this subsection, we will show that 
the sensitivity of KC atom interferometer will be greatly enhanced by quantum fluctuations of input MSS, and more importantly, this enhancement shows robustness in a sizable parameter region against strong Doppler effect by fully accounting for the imperfect atom-light interaction. 

\begin{figure}
	\includegraphics[width=1.0\linewidth]{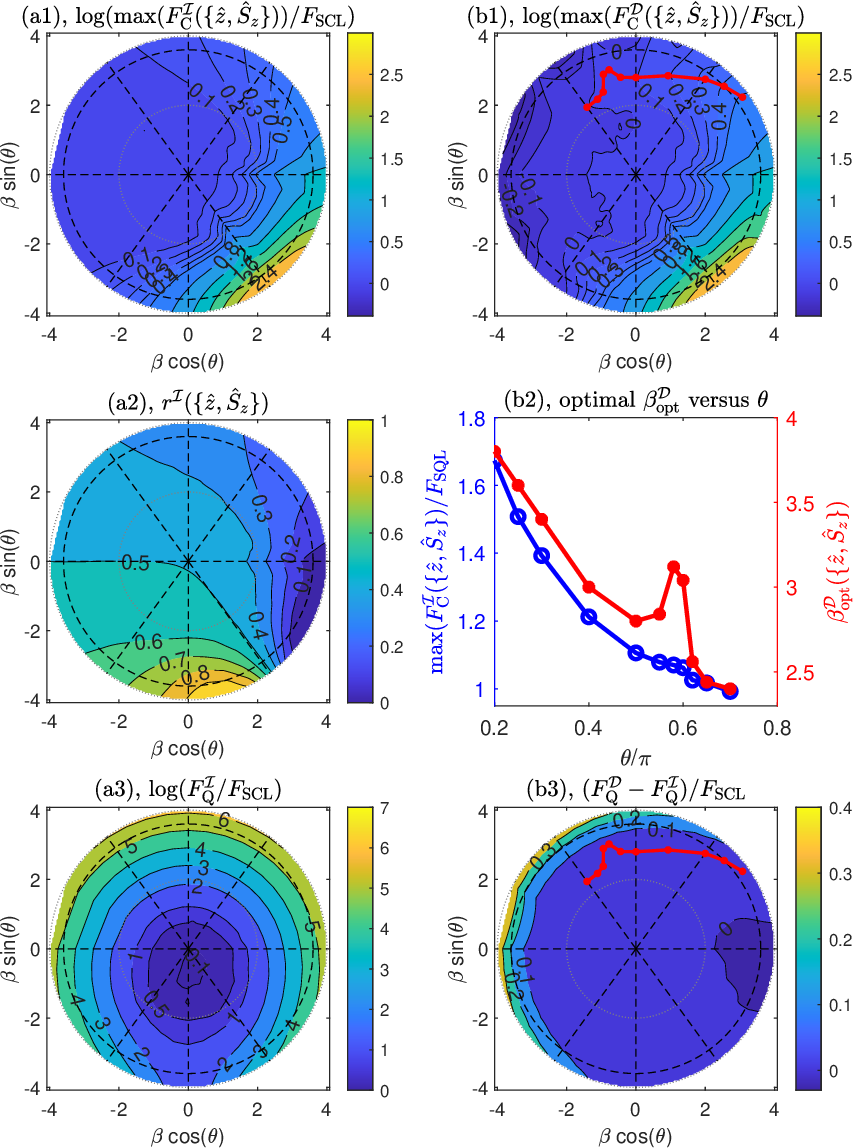}
	\caption{(color online) Evolution of max($F_{\rm C}^{\mathcal{S}}(\{\hat z,\hat S_z\})$) ((a1,b1), log scale), $r^{\mathcal{I}}(\{\hat z,\hat S_z\})$ (a2), and the corresponding QFI $F_{\rm Q}^{\mathcal{S}}$ (a3,b3) versus squeezing parameters $(\beta,\theta)$ for the "Ideal" ($\mathcal{S}=\mathcal{I}$) and "Doppler" ($\mathcal{S}=\mathcal{D}$) scenario. Particularly, $r^{\mathcal{D}}(\{\hat z,\hat S_z\})=r^{\mathcal{I}}(\{\hat z,\hat S_z\})$ is assumed for simplicity, as discussed in the main text. Panel (b2) shows the optimal squeezing $\beta^{\mathcal{D}}_{\rm opt}$ (the red solid line in panels (b1,b2,b3)) versus $\theta$ ($0\leq\theta\leq\pi$ is chosen) under "Doppler" scenario. 
	The black dashed radial lines (concentric circle)  correspond to $\theta/\pi=(0,0.3,0.5,0.7,1.0,1.3,1.7,2)$ ($\beta=3.6$). 
	Other parameters are the same as in Fig.~\ref{fig:fig3rd-1}.}
	\label{fig:fig3rd-2}
\end{figure}

In the "Ideal" scenario, as in last subsection, 
the light-atom interaction is perfect, 
the spatial distribution of $|s\rangle$ in the frame of $z_0=\frac{gT^2}{2}+\frac{\hbar k_0T}{2m}$ can be expressed as follow, 
\begin{align}
	P_{a/b}\propto  e^{-\frac{z_s^2+4z^2}{4\tilde\Delta_z^{ 2}}}\big(\cosh{\big(\frac{z_s z}{\tilde\Delta_z^2}\big)}
	\pm \cos({\tilde k} z+2\varphi)\big),
	\label{eqn:Prob-Ideal-zSz}
\end{align}
here $z_s=\hbar k_0(T_1-T_2)/m$ denotes the spatial separation of the two output squeezed Gaussian wave packets with width, 
\begin{eqnarray}
	\tilde\Delta_z=\sqrt{\frac{\hbar}{m\tilde\omega}\Big((\tilde\omega {\tilde T})^2+1\Big)},
\end{eqnarray}
and the effective wave-number $\tilde k$ of interference pattern between them,
\begin{eqnarray}
	{\tilde k}=k_0\frac{|T_2-T_1|}{\tilde T}\frac{(\tilde\omega {\tilde T})^{2}}{1+(\tilde\omega {\tilde T})^{2}}.
\end{eqnarray}
In fact, the effect of MSS can be attributed to the effective shift of the wave-number $\tilde k$ and width $\tilde\Delta_z$, as seen below,
\begin{eqnarray}
	{\tilde T}&=&T+\frac{1}{\omega}\frac{2R\sin(\theta)}{1+R^2-2R\cos(\theta)},\label{eqn:tildeT}\\ 
	\frac{1}{\tilde\omega}&=&\frac{1}{\omega}\frac{1-R^2}{1+R^2-2R\cos\theta},\label{eqn:tildeomega}
\end{eqnarray}
$R=\tanh(\beta)$. 
As a comparison, the case of $\beta=0$ corresponds to $\tilde\Delta_z=\sqrt{\frac{\hbar}{m\omega}\Big((\omega T)^2+1\Big)}$ and ${\tilde k}=k_0\frac{|T_2-T_1|}{T}\frac{(\omega T)^{2}}{1+(\omega T)^{2}}$. 
Obviously, Eq.~\ref{eqn:tildeT} is \textit{asymmetric} about the real axis of Poincare disk, Im$(\alpha)=0$.  
To be concrete, both the width $\tilde\Delta_z$ and the wavelength $2\pi/{\tilde k}$ of interference pattern in the lower half-plane of the Poincare disk ($\theta\in(\pi,2\pi)$) are decreased, compared to the ones in upper half-plane ($\theta\in(0,\pi)$), therefore, the former sensitivity is generically larger than the later, 
see Fig.~\ref{fig:fig3rd-2}(a1). 
As a consequence of this asymmetry, the corresponding optimal ratio $r^{\mathcal{I}}(\{{\hat z}, {\hat S}_z\})$ (solution by maximizing Eq.~\ref{eqn:CFI-zSz} in "Ideal" scenario) is also \textit{asymmetric} about Im$(\alpha)=0$ (Fig.~\ref{fig:fig3rd-2}(a2)). 
Particularly, $r^{\mathcal{I}}(\{{\hat z}, {\hat S}_z\})$ in this case can be larger than $1/2$, compared with the case of population measurement. 
In one word, we find that both the maximum sensitivity of KC atom inteferometry and the relative optimal ratio are  \textit{asymmetric}, which still hold on in the Doppler scenario. 

Furthermore, compared with population measurement, the sensitivity of joint measurement of population-position is greatly enhanced in a much sizable regime under the "Ideal" scenario, as depicted in Fig.~\ref{fig:fig3rd-2}(a1,a2,a3).  
However, this large enhancement strongly depends on the direction $\theta$ and the value of $\beta$. Usually, the larger $\omega T$, the smaller area of parameter space in which $\max {F_{\rm C}^{\mathcal{I}}(\{{\hat z}, {\hat S}_z\})}\gg F_{\rm SCL}$, 
and the direction $\theta_{\rm max}$ that maximize the CFI as fixing $\beta$ is roughly alongside  $\theta_c+\pi$, i.e., $\omega T=(0.32,2.0,12.5)$ corresponds to $\theta_{\rm max}\approxeq(1.25,1.7,1.95)\pi$. 
As we can see, the region in $(\beta,\theta)$ space in which large sensitivity enhancement are observed usually includes not only great position fluctuations but also momentum ones (Fig.~\ref{fig:fig3rd-2}(a1)), the later will make the "Ideal" scenario inaccurate any more and call for the "Doppler" one. 
Analogy to the case of thermal Doppler broadening in finite temperature, a critical question arises, what amount of this large quantum enhancement still persists with competition of strong Doppler suppression?

However, both the presence of Doppler effects and the nonlinear phase from MSS (Eq.~\ref{eqn:Psi0}) makes the numerical computation of Eq.~\ref{eqn:CFI-zSz} much more challenging to achieve the demanded accuracy by a high efficient way (see Appendix.~\ref{appendix:Levin} for details). 
As a comparison, the numerical computation of Eq.~\ref{eqn:CFI-Doppler-Sz} is easily established even in a laptop computer because of the absence of nonlinear phase from MSS. As a mark, the corresponding optimal control parameters for joint measurement of population and position are denoted as $r^{\mathcal{D}}(\{{\hat z},{\hat S}_z\})$,   $\psi_{12}^{\mathcal{D}}(\{{\hat z},{\hat S}_z\})$ and  $\psi_{23}^{\mathcal{D}}(\{{\hat z},{\hat S}_z\})$. 
In this work, to clearly quantify the Doppler effect of MSS when establishing joint measurement of population and position, 
we do not attempt to optimize the control parameters for joint measurement of population and position in the "Doppler" scenario, 
instead, we directly adopt the optimal control parameters obtained in the "Ideal" scenario by globally optimizing $F_{\rm C}^{\mathcal{I}}(\{{\hat z},{\hat S}_z\})$ (based on Eq.~(\ref{eqn:Prob-Ideal-zSz})) and then evaluate the corresponding CFI (Eq.~\ref{eqn:CFI-zSz}) and QFI (Eq.~\ref{eqn:QFI-Doppler-0}). That is, we assume \textit{a priori} 
$r^{\mathcal{D}}(\{{\hat z},{\hat S}_z\})=r^{\mathcal{I}}(\{{\hat z},{\hat S}_z\})$,
$\psi_{12}^{\mathcal{D}}(\{{\hat z},{\hat S}_z\})=\Phi^{\mathcal{I}}(\{{\hat z},{\hat S}_z\})$, and $\psi_{23}^{\mathcal{D}}(\{{\hat z},{\hat S}_z\})=0$. 

Fig.~\ref{fig:fig3rd-2}(b1,b3) show the corresponding $F_{\rm C}^{\mathcal{D}}(\{{\hat z},{\hat S}_z\})$ and $F_{\rm Q}^{\mathcal{D}}$ under "Doppler" scenario. 
As we can see, the competition between quantum enhancement of MSS and Doppler suppression divides the parameter space ($\beta,\theta$) into three regimes roughly.
Here we take $\omega T=0.32$ (equivalently $\sigma_p/\hbar k_0=0.1$) as an example. 
When $0.7\pi\lesssim\theta\lesssim1.2\pi$, the Doppler suppression completely dominates, resulting in a similar behavior between the two measurement protocols (both max($F_{\rm C}^{\mathcal{D}}(\{\hat S_z\})$) and max($F_{\rm C}^{\mathcal{D}}(\{{\hat z},\hat S_z\})$) are reduced largely below $F_{\rm SCL}$). 
While the quantum enhancement of MSS greatly enlarges the parameter area in which $\max(F_{\rm C}^{\mathcal{D}}(\{{\hat z},\hat S_z\}))\gg F_{\rm SCL}$, here $-0.6\pi \lesssim \theta \lesssim 0.2\pi$, this is sharply distinct from the population measurement (Fig.~\ref{fig:fig3rd-1}(b1)). 
Furthermore, besides these regimes, the competition also induces an optimal squeezing parameter $\beta^{\mathcal{D}}_{\rm opt}$ where $\max(F_{\rm C}^{\mathcal{D}}(\{{\hat z},\hat S_z\}))$ reaches the maximum by varying $\beta$ and fixing $\theta$, see the red solid lines in Fig.~\ref{fig:fig3rd-2} (b1,b2,b3).  We note here that it is only shown that the optimal squeezing $\beta^{\mathcal{D}}_{\rm opt}$ versus $\theta$ in the upper plane. 
A quantitative comparison between the measurement protocols under both scenarios is left in Appendix.~\ref{appendix:Comparison}. 

\begin{figure}
	\centering
	\includegraphics[width=1.0\linewidth]{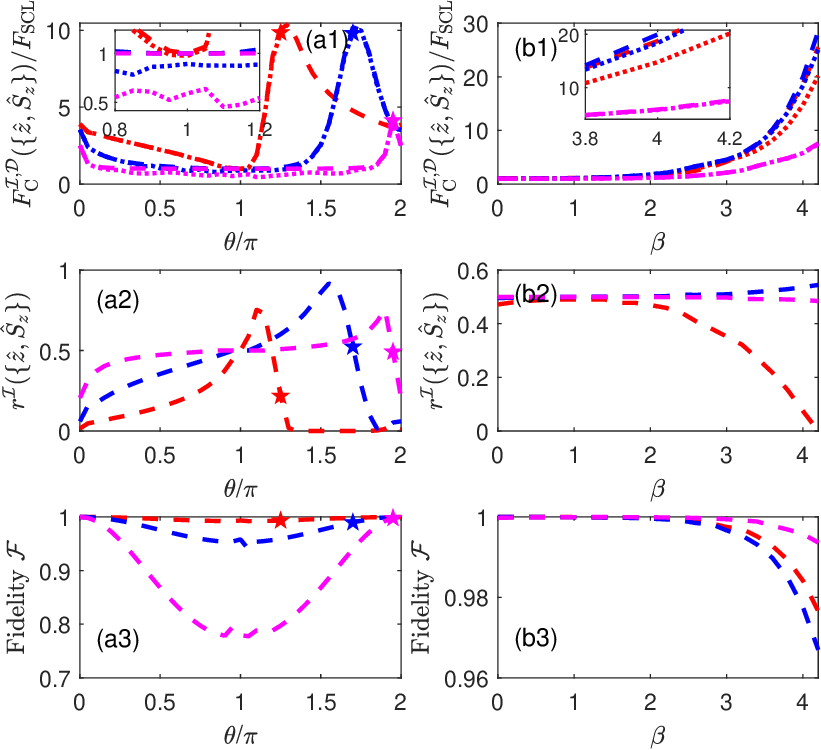}
	\caption{(color online) Under the joint measurement of population-position, max($F_{\rm C}^{\mathcal{I,D}}(\{\hat z, \hat S_z\}))$) (a1, b1), optimal control ratio $r^{\mathcal{I}}(\{\hat z, \hat S_z\}))$ (a2, b2), and the fidelity $\mathcal{F}$ (a3, b3) are shown as functions of the squeezing parameters $\theta$ (with $\beta = 3.6$ fixed, panels (a1, a2, a3), see the black dashed circles in Fig.~\ref{fig:fig3rd-2}) and $\beta$ (with $\theta = \theta_{\rm max}$ fixed, panels (b1, b2, b3)). Here, the red, blue, and magenta dashed curves (pentagram symbols) correspond to $T\omega=0.32$ ($\beta=3.6,\theta=\theta_{\rm max}=1.25\pi$), $2.0$ ($\beta=3.6,\theta=\theta_{\rm max}=1.7\pi$), and $12.5$ ($\beta=3.6,\theta=\theta_{\rm max}=1.95\pi$), respectively. In panels (a1,b1), the dashed (dotted) lines denote the "Ideal" ("Doppler") scenario. Other parameters are the same as in Fig.~\ref{fig:fig3rd-1}.}
	\label{fig:fig2nd-5}
\end{figure}

The joint measurement of population and spatial distribution $\{\hat z, \hat S_z\}$ is of significant scientific and practical value in real experiments. 
To further illustrate the impact of this measurement and the role of Doppler effects on sensitivity enhancement—and to more clearly demonstrate the potential applications of motional squeezed states, 
we present in Fig.~(\ref{fig:fig2nd-5}) the maximum $F_{\rm C}^{\mathcal{I,D}}(\{\hat z,\hat S_z\})$, the corresponding optimal ratio $r^{\mathcal{I}}(\{\hat z,\hat S_z\})$, and the associated fidelity $\mathcal{F}$ (Eq.~(\ref{eqn:fidelity})) for this measurement under both the “Ideal” and “Doppler” scenarios. 
Specifically, we either fix $\beta = 3.6$ and scan the angle $\theta \in [0, 2\pi]$ (Fig.~\ref{fig:fig2nd-5}(a1,a2,a3)), or fix $\theta = \theta_{\rm max}$ (where $\theta_{\rm max}$ is defined as the angle at which $F_{\rm C}^{\mathcal{I}}(\{{\hat z,\hat S_z}\})$ reaches the maximum for a fixed $\beta$) and scan $\beta \in [0, 4.2]$ (Fig.~\ref{fig:fig2nd-5}(b1,b2,b3)). 
As we can see, the amplitude of CFI can be roughly magnified by one order around $\theta=\theta_{\rm max}$ for large enough $\beta$ (see the pentagrams in Fig.~\ref{fig:fig2nd-5}), and the Doppler suppression is weak (Fig.~\ref{fig:fig2nd-5} (a1,b1)). However, the strong Doppler effects do not only suppress the CFI but also decrease the fidelity non-negligibly (Fig.~\ref{fig:fig2nd-5} (a3,b3)), the later would be a matter for the practical deployment of KC atom interferometer with MSSs. 
We note here that $F_{\rm C}^{\mathcal{D}}(\{\hat z,\hat S_z\})$
was not computed when the corresponding $r^{\mathcal{I}}(\{\hat z,\hat S_z\})$ approaches zero (Fig.~\ref{fig:fig2nd-5} (a1)) because the numerical integral in Eq.~\ref{eqn:CFI-zSz} becomes unstable there.

\section{Conclusion and Outlook}
\label{sec4}
By utilizing the unitary evolution operator describing the light–atom interaction under arbitrary momentum broadening, 
we supply a generic computational framework of quantum and classical Fisher information of two commonly-deployed measurement protocols by fully accounting for the Doppler effects. 
Further, this generic framework enables us to quantitatively analyze the impact of external quantum fluctuation and Doppler suppression of Kasevich-Chu atom interferometer by inputting a  single-mode MSS. 

Then we employ global optimization algorithm to maximize the CFI of each measurement both in the "Ideal" and "Doppler" scenario to determine the optimal control parameters and the corresponding QFI. 
For population measurement, we find that the sensitivity can be enlarged as four times the SCL 
around a small area where position fluctuation dominates over momentum one. 
While for joint measurement of population and position, 
In case of ignoring Doppler effect of momentum fluctuation, 
the impact of MSS can be mapped into the re-normalization of output spatial spread and wavelength of intereference pattern, which not only induces asymmetry in Poincare disk of MSS but also shows great quantum enhancement of sensitivity. 
More importantly, in fully accounting for the Doppler effects, 
the competition between quantum enhancement and Doppler suppression divides the squeezing parameter space into three regimes typically. 
In one regime where quantum enhancement dominates over Doppler suppression, the ultimate sensitivity can be enhanced by more than one order of SCL, even under standard-quantum limit; 
while for another regime, an optimal squeezing parameter $\beta^{\mathcal{D}}_{\rm opt}$ is induced from these two competing mechanism. 

As stated in previous sections, the large momentum fluctuation of MSS will greatly degrade the fidelity between "Ideal" and "Doppler" scenario, that is to say, the interaction between pulsed light and atom is imperfect, which decreases the total atom number that effective interferes in the interferometer and therefore suppress the final sensitivity. 
There are several possible approaches to mitigate the imperfect pulses caused by strong Doppler effects. For example, one may employ shaped or modulated pulse schemes~\cite{YukunLuo2016}, or Floquet engineering to realize perfect $\pi/2$ and $\pi$ pulses~\cite{Wilkason2022}. 
Multi-path interference scheme~\cite{WangYiping2024} is also a promising approach to ultimately enhance the sensitivity of atom interferometers with strong Doppler effects. 

Additionally, In a traditional Kasevich-Chu atom gravimeter based on cold alkali-metal atomic gas, 
the light-atom interaction is usually engineered by Raman scattering, 
where the decoherence of the third electronic excited state may take a significant role because
the large momentum fluctuation should reduce the effective single-photon detuning. 
Therefore, our proposal would be better applied to the interferometry based on cold alkaline-earth(-like) atomic gas  where the two-level of atom is directly coupled by a single laser and the natural width of electronic-excited state is much narrow~\cite{ZhangXian2016}. 

Finally, our findings clearly demonstrate the robustness of quantum enhancement of MSS under strong Doppler suppression, 
and hold significant potential for an unexplored external degree of freedom, especially when integrated into a mobile Kasevich‑Chu atom interferometer~\cite{Parazzoli2012}. 
Our study can also be extended to atom interferometers based on spin-squeezing of multi-component Bose–Einstein condensates~\cite{Szigeti2020}. 

\section{Acknowledgment}
We thanks stimulating discussions on motional squeezing states and SU(2) Lie group theory with Chenwei Lv and Zhan Zheng, and helpful details of Kasevich-Chu atom interferometry input from Yin Zhou and Kanxing Weng. 
We thanks helpful discussions on the numerical integral on Eq.~\ref{eqn:CFI-zSz} and computation supports from Dr.~Xitao Yu. 
We also thanks helpful suggestions on organization of our manuscript from Wenqiang Zheng. And we thanks Chenwei Lv whom reads our manuscript thoroughly and gives the corresponding many suggestions. 

\appendix

\section{Quantum Fisher Information for arbitary momentum broadening}
\label{appendix:QFI-Doppler}
Under the chirping rate equivalent to $k_0gt$, the unitary operators ${\hat U}^{\mathcal{I},\mathcal{D}}_j$ are independent on the gradient of gravitational potential $g$ whatever the scenario is. This feature makes possible to calculate the quantum Fisher information (QFI) even in the presence of strong Doppler effect. All the calculation are  established with help of the software  \textit{Mathematica} and the package \textit{DiracQ}~\cite{Wright2015} which help us to collect, extract and simplify all the coefficients as stated below. According to the definition of QFI, $\frac{F_Q}{4}=\langle \partial_g\Psi(t)|\partial_g\Psi(t)\rangle-|\langle\Psi(t)|\partial_g\Psi(t)\rangle|^2$, the QFI for arbitrary single-particle input state $|\Psi_0\rangle$ (under "Doppler" scenario) can be written down as follow,
\begin{eqnarray}
	\frac{F_Q^{\mathcal{D}}}{4}&=&\langle\Psi_0|{\hat c}_0+{\hat c}_1{\hat z}+{\hat c}_2{\hat z}^2|\Psi_0\rangle\nonumber\\
	&&-|\langle\Psi_0|{\hat d}_0+d_1{\hat z}|\rangle\Psi_0\rangle|^2,
\end{eqnarray}
where both $c_2=d_1^2$ and $d_1=mT/\hbar$ are constants, while ${\hat c}_{0}$, ${\hat c}_{1}$ and ${\hat d}_0$ solely depend on the momentum operator. 
The later can be further expanded as a summary around $g=0$, 
\begin{eqnarray}
	{\hat c}_0&=&\sum_{j=0}^2 {\hat c}_{0,j}g^j,\nonumber\\
	{\hat c}_1&=&\sum_{j=0}^1 {\hat c}_{1,j}g^j,\nonumber\\
	{\hat d}_0&=&\sum_{j=0}^1 {\hat d}_{0,j}g^j.
\end{eqnarray}
By fully employing the equivalences as stated below,
\begin{eqnarray}
	{\hat c}_{0,2}&=&d_{0,1}^2,\nonumber\\
	{\hat c}_{0,1}&=&2{\hat d}_{0,0}d_{0,1},\nonumber\\
	{\hat c}_{1,1}&=&2d_1d_{0,1},
\end{eqnarray}
the QFI in the "Doppler" scenario can be recast as follow, 
\begin{eqnarray}
	\frac{F_Q^{\mathcal{D}}}{4}&=&\langle\Psi_0|{\hat c}_{0,0}+{\hat c}_{1,0}{\hat z}+{\hat c}_2{\hat z}^2|\Psi_0\rangle\nonumber\\
	&&-|\langle\Psi_0|{\hat d}_{0,0}+d_1{\hat z}|\rangle\Psi_0\rangle|^2.
\end{eqnarray}
It would be heuristic to reformulate the QFI in a similar manner as in the "Ideal" scenario by introducing ${\hat G}^\prime$ and $\hat G^{\prime\prime}$, 
\begin{eqnarray}
	F_Q^{\mathcal{D}}&=&k_0^2\big(\frac{T^2}{2}-T_2^2\big)^2\nonumber\\
	&&+4\langle\Psi_0| \hat G^{\prime\prime}|\Psi_0\rangle+4{\rm Var}({\hat G}^\prime),\nonumber\\
\end{eqnarray}
here
\begin{eqnarray}
	{\hat G}^\prime=\hat d_{0,0}+d_1{\hat z},
\end{eqnarray}
and 
\begin{align}
	{\hat G}^{\prime\prime}={\hat c}_{0,0}+{\hat c}_{1,0}{\hat z}+c_{2}{\hat z}^2-{\hat G}^{^\prime 2}-\frac{k_0^2}{16}\big(T^2-2T_2^2\big)^2,\nonumber\\
	= {\hat c}_{0,0}-{\hat d}_{0,0}^2-i\hbar d_1 \partial_p{\hat d}_{0,0}
	-\frac{k_0^2}{16}\big(T^2-2T_2^2\big)^2,\nonumber\\
	=\sum_{j=4}^2 \sum_{i=0}^{4} {\hat G}^{\prime\prime}_{i,j-i}T_1^{i}T_2^{j-i},
\end{align}
in the later case, ${\hat G}^{\prime\prime}$ only depends on the momentum, while the dependence on position operator ${\hat z}$ is eliminated by employing the equivalences above. 

\begin{figure}
	\includegraphics[width=0.5\textwidth]{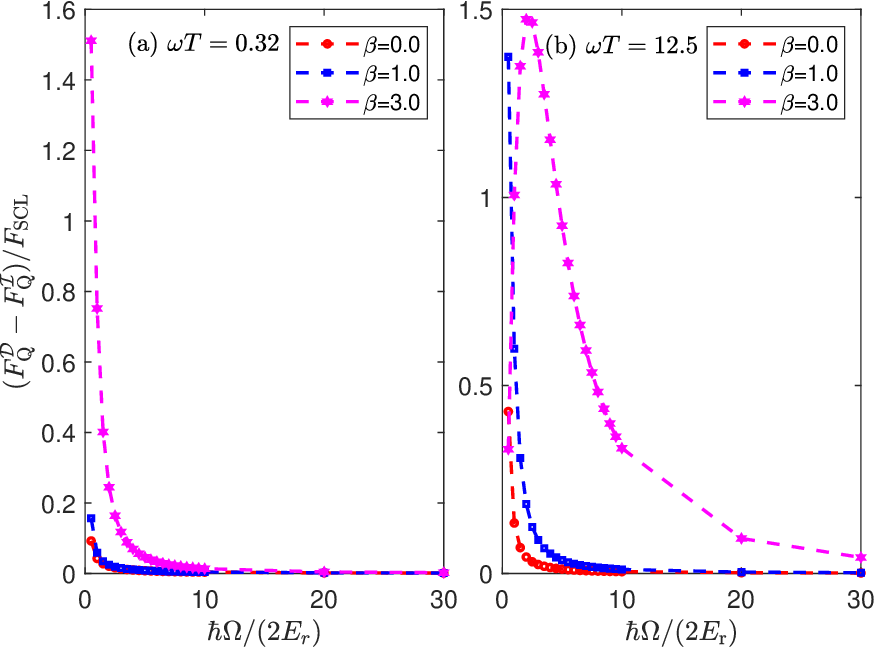}
	\caption{(color online) Doppler shifts of QFI difference $\Delta F_{\rm Q}=F_{\rm Q}^{\mathcal{D}}-F_{\rm Q}^{\mathcal{I}}$ versus Rabi frequency $\Omega$ as $\omega T=0.32$ (a) and $\omega T=12.5$ (b). 
	Here we set $r=T_1/T=1/2$ artificially. 
	For the input MSS, $\theta=\pi$ is chosen where the Doppler broadening is the largest and $\beta=(0,1,3)$ (dashed red, blue and magenta lines). Other parameters are the same as in Fig.~\ref{fig:fig3rd-1}.}
	\label{fig6}
\end{figure}

Usually, ${\hat G}^{\prime\prime}$ can be further approximated if $k_0^2T$ is large enough,  i.e., 
the coefficients $|G^{\prime\prime}_{i,3-i}|$ ($i=1,\cdots,3$),  
\begin{eqnarray}
	G^{\prime\prime}&\approx&\sum_{i=0}^{4} {\hat G}^{\prime\prime}_{i,4-i}T_1^{i}T_2^{4-i},
\end{eqnarray}
\begin{eqnarray}
	{\hat G}^{\prime\prime}_{0,4}=-k_0^2\left( (\frac{1}{2} - {\hat C}_1^2)(\frac{1}{2} - {\hat C}_2^2) -  \frac{\hat\zeta}{4} \right)^2,\\
	{\hat G}^{\prime\prime}_{1,3}=\frac{k_0^2}{2}\Big( \hat\zeta^\prime({\hat C}_2^2 - \frac{1}{2}) - ({\hat C}_1^2 - \frac{1}{2}) \hat\zeta +\frac{1}{2}\Big),\\
	{\hat G}^{\prime\prime}_{2,2}=\frac{k_0^2}{4}\Big(   \hat\zeta^\prime({\hat C}_2^2 - \frac{3}{2}) - ({\hat C}_1^2 - \frac{1}{2})\hat\zeta -\frac{1}{2}\Big),\\
	{\hat G}^{\prime\prime}_{3,1}=-k_0^2 (\frac{1}{2} -  {\hat C}_1^2)^2,\\
	{\hat G}^{\prime\prime}_{4,0}=-\frac{k_0^2}{4} (\frac{1}{2} -  {\hat C}_1^2)^2,
\end{eqnarray}
$\hat\zeta=4{\hat C}_1{\hat C}_2\operatorname{Re}(( {\hat A}_1 + i{\hat B}_1) ( {\hat A}_2 + i{\hat B}_2) e^{i\hat\Theta})$ (Eq.~\ref{eqn:Theta}), $\hat\zeta^\prime=4{\hat C}_1^2({\hat C}_1^2 - 1)$.
In the "Ideal" scenario ($\Omega\rightarrow\infty$), 
\begin{eqnarray}
	\label{eqn:IdealConditions}
	\begin{array}{|c|c|c|}
		\hline
		{\hat A}_1=\frac{1}{\sqrt{2}} & {\hat B}_1=0 & {\hat C}_1=\frac{1}{\sqrt{2}} \\
		\hline 
		{\hat A}_2=0 & {\hat B}_2 =0 & {\hat C}_2 =1,\\
		\hline 
		{\hat A}_3=\frac{1}{\sqrt{2}} & {\hat B}_3=0 & {\hat C}_3=\frac{1}{\sqrt{2}}\\
		\hline
	\end{array}
\end{eqnarray}
as a result, ${\hat G}^{\prime\prime}_{i,4-i}=0$ for $i=0,\ \cdots,\ 4$. Actually, ${\hat G}^{\prime\prime}_{i,j}=0$ exactly holds on in the "Ideal" scenario. 
A similar approximation on ${\hat d}_{0,0}$ also can be written down as follow, 
\begin{eqnarray}
	{\hat d}_{0,0}&=&\sum_{j=2}^1\sum_{i=0}^2 {\tilde{d}}_{i,j-i}T_1^{i}T_2^{j-i}\nonumber\\
	&&\approx\sum_{i=0}^2 {\tilde{d}}_{i,2-i}T_1^{i}T_2^{2-i},
\end{eqnarray}
here 
\begin{align}
	{\tilde{d}}_{2,0}=\frac{k_0}{2}({\hat C}_1^2 + \frac{{\hat p}}{\hbar k_0}),\nonumber\\
	{\tilde{d}}_{1,1}=k_0({\hat C}_1^2  + \frac{{\hat p}}{\hbar k_0}),\nonumber\\
	{\tilde{d}}_{0,2}=\frac{\hat p}{2\hbar}+\frac{k_0}{2} \Big( {\hat C}_1^2+{\hat C}_2^2-2{\hat C}_1^2{\hat C}_2^2 + \frac{\hat\zeta}{2} \Big).
\end{align}
In the "Ideal" scenario, $2{\tilde d}_{2,0}=2{\tilde d}_{0,2}={\tilde d}_{1,1}=k_0(\frac{1}{2}  + \frac{{\hat p}}{\hbar k_0})$.
As a result, ${\hat d}_{0,0}=\frac{k_0T^2}{4}+\frac{T^2}{2\hbar}{\hat p}$, 
which proves that the coincidence on QFI between the "Ideal" and "Doppler" scenario, that is,
\begin{eqnarray}
	\lim_{\Omega\rightarrow\infty}{\hat G}^{\prime\prime}=0,\\
	\lim_{\Omega\rightarrow\infty}{\hat G}^\prime={\hat G}+\frac{k_0T^2}{4}.
\end{eqnarray}
as depicted in Fig.~\ref{fig6}


\section{Numerical caluculation of $F_{\rm C}^{\mathcal{D}}(\{{\hat z},{\hat S}_z\})$}
\label{appendix:Levin}
In this appendix, we add the details on numerical calcuation of $F_{C}^{\mathcal{D}}(\{{\hat z},{\hat S}_z\})$ with the assistence of Levin-type quadrature. 

According to the definitions of ${\hat K}_{sa}$ ${\hat K}_{sa}^\prime$ and ${\hat K}_{sa}^{\prime\prime}$, both $L_s(z)$ and $H_s(z)$ can be transformed into a uniform formula, $D(z)=\int dp \frac{K(p)}{\sqrt{2\pi\hbar}}e^{-(iA+B)p^2+ipz/\hbar}$ ($B>0$), where $A$ and $B$ depend on the squeezing parameters $(\beta,\theta)$, 
\begin{eqnarray}
	A=\frac{1}{2m\hbar}\Big(\frac{2R\sin(\theta)}{1+R^2-2R\cos\theta}\frac{1}{\omega}+T\Big),
\end{eqnarray}
and
\begin{eqnarray}
	B=\frac{1}{2m\hbar}\frac{1-R^2}{1+R^2-2R\cos(\theta)}\frac{1}{\omega}.
\end{eqnarray}
While the kernel $K(p)$ can be finally expanded as follow, 
\begin{eqnarray}
	K(p)=\sum_{j=1}^{N_0} \frac{a_{j,0}a_{j,1}(p)}{\sqrt{2\pi\hbar}}\exp\left(i{\tilde b}_jp+ic_j\bar\Delta(p)\right),
\end{eqnarray}
where $\bar\Delta(p)=\sqrt{1+P^2}$,  $P=\frac{k_0}{m\Omega}p+\bar\delta$,   $\bar\delta=\frac{\delta_0+E_r/\hbar}{\Omega}$, and 
${\tilde b}_j=(b_j+z/\hbar)$ depends on the position $z$. And all coefficients $a_{j,0}$, $b_j$ and $c_j$ are independent on the momentum and already obtained analytically while $a_{j,1}(p)$ belongs to the set (seven functions on the momentum),  $(P\pm\bar\Delta(p))^l/\bar\Delta(p)^3$ ($l=0,1,2,3$). 
The number $N_0$ depend on the choice of ${\hat K}_{sa}$, ${\hat K}_{sa}^\prime$ or ${\hat K}_{sa}^{\prime\prime}$.  
This integral on $D(z)$ exhibits three key features: 
(i) a highly oscillatory factor, $\exp(-iAp^2)$ where the phase is quadratic on $p$;
(ii) a momentum-dependent nonlinear phase, $\exp(i c_j \bar{\Delta}(p))$;
(iii) an analytic continuation of $a_{j,1}(p)$ into the complex plane that spans multiple Riemann sheets.
The second feature (ii) precludes the use of Gaussian integration to obtain a closed-form expression, while the third (iii) renders analytic continuation into the complex plane prohibitively complicated, preventing any straightforward contour deformation to suppress the oscillatory term. Moreover, due to the first feature (i), standard Fourier transform techniques may also break down in the regime $|A| \gg B$. 

To handle the classical Fisher information for MSSs under joint measurement of number and postion, especially at large momentum broadening, within a unified framework, we employ a hybrid scheme combining domain decomposition and Levin-type quadrature~\cite{Levin1996,Levin1997}. First, we divide the whole integral into two domains according to the sign of $P$ which gives  $p_0=-m(\delta_0+E_r/\hbar)/k_0$, as a result, 
\begin{eqnarray}
	D=\sum_{j=1}^{N_0}\int_{-\infty}^{p_0}dp e^{i\phi_L(p)}f_L(p)
	+\int_{p_0}^{\infty}dp e^{i\phi_R(p)}f_R(p),
\end{eqnarray}
and 
\begin{align}
	f_L(p)=\frac{a_{j,0}a_{j,1}(p)}{\sqrt{2\pi\hbar}}e^{-Bp^2}e^{ic_j(\bar\Delta(p)+P)},\nonumber\\
	f_R(p)=\frac{a_{j,0}a_{j,1}(p)}{\sqrt{2\pi\hbar}}e^{-Bp^2}e^{ic_j(\bar\Delta(p)-P)}.
\end{align}
In this formula, the kernels $f_{\rm L,R}(p)$ is a smooth function on $p$ which does not include any highly-oscillatory term due to $\lim_{p\rightarrow-\infty}\bar\Delta(p)+P=0$ and $\lim_{p\rightarrow\infty}\bar\Delta(p)+P=0$. 
While the phases $\phi_{\rm L,R}(p)$ are now at most quadratic polynomials in the momentum $p$, with no additional nonlinear dependence on $p$, 
\begin{align}
	\phi_L(p)=-c_j\bar\delta+(b_j+\frac{z}{\hbar}-\frac{k_0}{m\Omega} c_j)p-Ap^2,\nonumber\\	
	\phi_R(p)=+c_j\bar\delta+(b_j+\frac{z}{\hbar}+\frac{k_0}{m\Omega} c_j)p-Ap^2.
\end{align}
Accordingly, two stationary-points are defined as follows, 
$\frac{d\phi_L(p)}{dp}|_{p=p_{L,*}}=0$ and $\frac{d\phi_R(p)}{dp}|_{p=p_{R,*}}=0$, with 
\begin{eqnarray}
	p_{L,*}=\frac{b_j+\frac{z}{\hbar}-\frac{k_0}{m\Omega} c_j}{2A},\ 
	p_{R,*}=\frac{b_j+\frac{z}{\hbar}+\frac{k_0}{m\Omega} c_j}{2A}.
\end{eqnarray}
Finally, this numerical integral on $D(z)$ can be handled within a unified framework. Specifically, near each stationary-point, the phase is locally linear in momentum~\cite{Lovetskiy2024}, allowing for highly efficient and accurate evaluation via Gauss–Legendre quadrature; while away from the stationary-point, Levin-type methods, which are not effective near stationary-points, enable accurate  computation. 

\begin{figure}
	\includegraphics[width=1.0\linewidth]{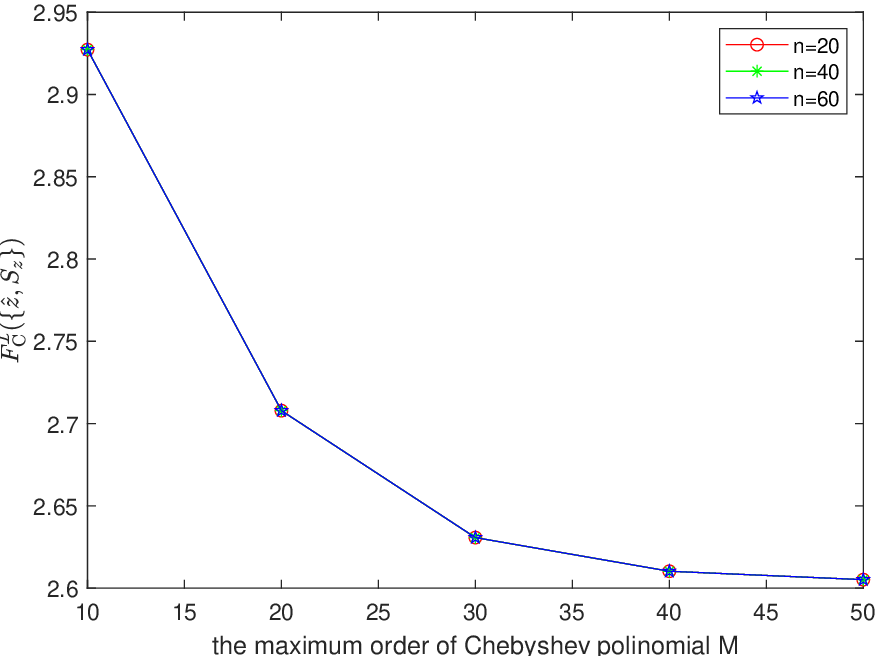}
	\caption{(color online) Benchmark of $F_{\rm C}^{\mathcal{D}}(\{\hat z, {\hat S}_z\})$ versus the maximum order of Chebyshev polynomials $M$ and the maximum order of Gauss-Legendre quadrature $n$. Here $\beta=4.6$ and $\theta=3\pi/2$. Other parameters are the same as in the main text.}
	\label{fig7}
\end{figure}

In the following, we simply introduce the Levin-type quadrature, 
\begin{eqnarray}
	{\mathcal{L}}=\int_{p_d}^{p_u}dpe^{i\phi(p)}f(p),
\end{eqnarray}
where the stationary-point of phase $\phi(p)$ is supposed to be far away from the interval $(p_d,p_u)$. 
This integral can be obtained by 
\begin{eqnarray}
	\mathcal{L}=F(p_u)e^{i\phi(p_u)}-F(p_d)e^{i\phi(p_d)},
\end{eqnarray}
provided that the first-order differential equation on $F(p)$ can be obtained numerically~\cite{Levin1996}, 
\begin{eqnarray}
	\frac{dF(p)}{dp}+i\frac{d\phi(p)}{dp}F(p)=f(p).
\end{eqnarray}
The differential equation on $F(p)$ can be solved numerically by expanding it with truncated Chebyshev polynomials as follow, 
\begin{eqnarray}
	F(p)=\sum_{j=0}^M {\tilde c}_jT_j(x),
\end{eqnarray}
here $p=\frac{p_u+p_d}{2}+\frac{p_u-p_d}{2}x$ and $T_j(x)$ is the $j$th Chebyshevy polynomial. Usually, Chenshaw-Curtis nodes are chosen~\cite{Iserles2025}, $x_k=-\cos(\pi\frac{k}{N})$, $k=0,\cdots, N$. Therefore, 
\begin{eqnarray}
	\sum_{j=0}^M {\tilde c}_j\Big(T_j^\prime(x_k)+i\tilde\phi^\prime(x_k)T_j(x_k)\Big)=\frac{p_u-p_d}{2}\tilde{f}(x_k),
\end{eqnarray}
here $T_j^\prime(x)$ is the first-order derivative of $T_j(x)$ with respect to $x$, $\tilde\phi^\prime(x)\equiv\frac{d\phi(p)}{dx}$, $\tilde f(x)\equiv f(p)$. It is worthy noting that $N\approx3M$ is usually chosen to balance the accuracy and efficiency. 
Finally, the coeficients ${\tilde c}_j$ and the numerical value $F(p)$ at both ends $p_u$ and $p_d$ can be obtained 
by least-squares fit, because this equation is overdetermined $N\gg M$. In Fig.~\ref{fig7}, we demonstrate the benchmark of $F_{\rm C}^{\mathcal{D}}(\{\hat z, {\hat S}_z\})$ versus versus the maximum order of Chebyshev polynomials $M$ and the maximum order of Gauss-Legendre quadrature $n$. As we can see, $F_{\rm C}^{\mathcal{D}}(\{\hat z, {\hat S}_z\})$ saturates around $M=40$, while the maximum order of  Gauss-Legendre quadrature $n$ has negligible effect. Thorough this work, we implement numerical integral of Eq.~\ref{eqn:CFI-zSz} with  $M=50$ and $n=60$. 


\section{Comparison the population measurement and the joint measurement of population-position in both scenarios}
\label{appendix:Comparison}
\begin{figure*}[t]
	\centering
	\includegraphics[width=1.0\linewidth]{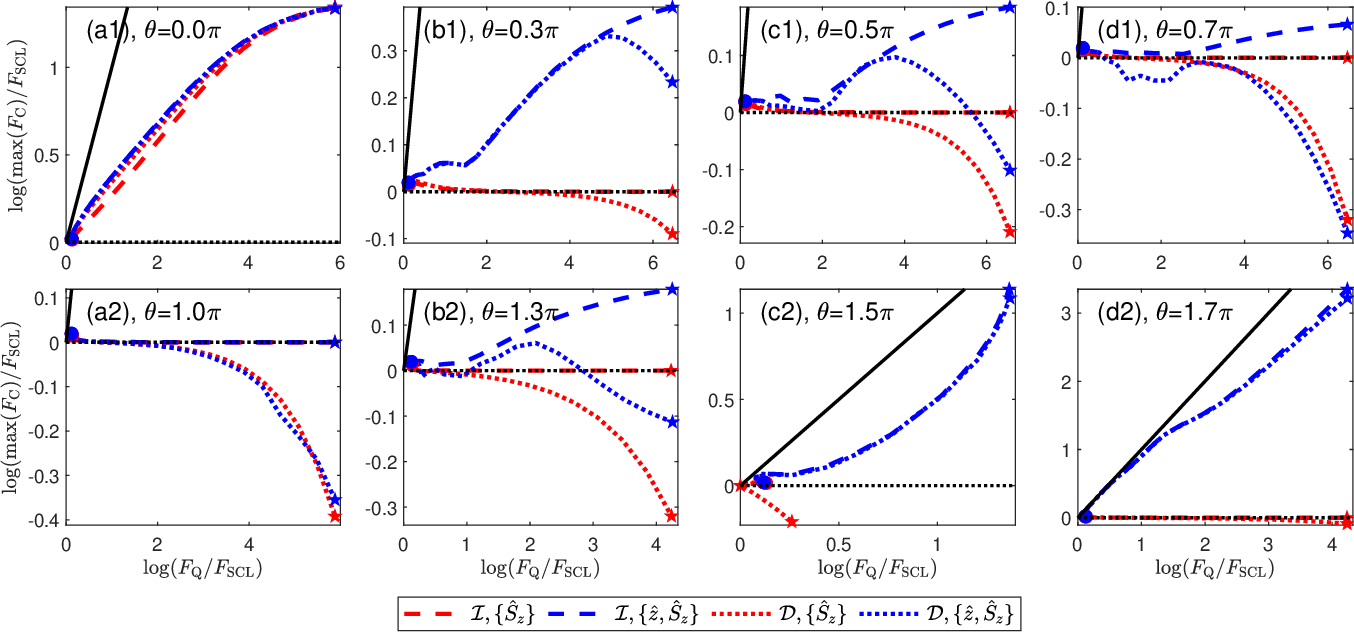}
	\caption{(Color online) Maximum CFIs (vertical axis, log scale) versus the corresponding QFIs (horizontal axis, log scale) for two measurement protocols and two scenarios along four diagonal directions (see the black dashed radial lines in Fig.~\ref{fig:fig3rd-1} and Fig.~\ref{fig:fig3rd-2}), with $\beta$ varying from $0$ (filled circles) to $4.2$ (pentagrams), the explicit dependence of all CFIs and QFIs on the squeezing parameter $\beta$ is suppressed in each panel. Dashed (dotted) curves correspond to the “Ideal” (“Doppler”) scenario, while red (blue) lines denote population measurement $\{\hat S_z\}$ (joint measurement of population-position $\{\hat z,\hat S_z\}$). The black dotted and solid lines indicate the SCL and QCRB, respectively. Other parameters are identical to those in Fig.~\ref{fig:fig3rd-1}.}
	\label{fig:fig2nd-4}
\end{figure*}

To simultaneously demonstrate the quantum enhancement and the corresponding Doppler suppression of sensitivity for two measurement protocols, 
here we choose four typical diagonal directions in the Poincare disk (the black dashed radial lines in Fig.~\ref{fig:fig3rd-1} and Fig.~\ref{fig:fig3rd-2}) while scanning $\beta$ from zero to a finite value (here max($\beta$)$=4.2$) and suppress the explicit dependence on the squeezing parameter $\beta$ in every panel to clearly demonstrate their QCRBs, see Fig.~\ref{fig:fig2nd-4} as a comprehensive comparison. 

Around $\theta=0$, both momentum fluctuation and Doppler effect are negligible, then the differences between scenarios are minimal, and the atom coherence length is so large that the measurement gain of two protocols also merges as increasing $\beta$, see Fig.~\ref{fig:fig2nd-4}(a1). 
However, around $\theta=\pi$, 
the strongest momentum fluctuations not only completely destroy the quantum enhancement in the "Ideal" scenario ($F_{\rm C}^{\mathcal{I}}(\{\hat S_z\})\approx F_{\rm C}^{\mathcal{I}}(\{\hat z, \hat S_z\})\approx F_{\rm SCL}$, see the red lines ), but also induce large Doppler effects which further suppress the remained measurement gain below the SCL ($F_{\rm C}^{\mathcal{D}}(\{\hat S_z\})\approx F_{\rm C}^{\mathcal{D}}(\{\hat z, \hat S_z\}) \ll F_{\rm SCL}$), see Fig.~\ref{fig:fig2nd-4}(a2). 

As squeezing parameter $\theta$ moving away from these two points, the difference between two measurement protocols demonstrates themselves explicitly, see Fig.~\ref{fig:fig2nd-4}. 
In the "Ideal" scenario, the MSS only introduces a negligible enhancement under population measurement (red dashed lines in Fig.~\ref{fig:fig2nd-4}), 
while brings much larger promotion under joint measurement of population-position (blue dashed lines in Fig.~\ref{fig:fig2nd-4}). 
In the "Doppler" scenario where the Doppler effect is fully taken into consideration, 
although the Doppler effects suppress the sensitivity in both measurement protocols, 
the joint measurement of population-position shows much better robustness against the Doppler suppression, compared with the population measurement. 
Eventually the quantum enhancement of joint measurement of population-position can completely dominate the Doppler suppression (Fig.~\ref{fig:fig2nd-4} (c2,d2)) which is absent in the population measurement. 

\bibliography{mybib}
\end{document}